\documentstyle[prl,eqsecnum,aps]{revtex}

\begin{document}
\author{Shen Jian-Qi $^{1,2}$ \footnote{E-mail address: jqshen@coer.zju.edu.cn}$,$ Xiao San-Shui $^{1}$  and Wu Qiang $^{2}$}
\address{1. Centre for Optical and Electromagnetic Research, State Key Laboratory of Modern
Optical Instrumentation, College of Information Science and
Engineering   \\ 2. Department of Physics and Zhejiang Institute
of Modern Physics   \\ Zhejiang University (YuQuan), Hangzhou
310027, China }

\date{\today}

\title{ Exact expression for decoherence factor in the
time-dependent\\ generalized Cini model \footnote{ This project is
supported in part by the National Natural Science Foundation of
China under the project No. $90101024$.}}

\maketitle

\begin{abstract}
The present letter finds the complete set of exact solutions of
the time-dependent generalized Cini model by making use of the
Lewis-Riesenfeld invariant theory and the invariant-related
unitary transformation formulation and, based on this, the general
explicit expression for the decoherence factor is therefore
obtained. This study provides us with a useful method to consider
the geometric phase and topological properties in the
time-dependent quantum decoherence process.

OCIS: 270.5570, 270.0270, 270.1670

\end{abstract}


Quantum decoherence problem is one of the most important and
interesting topics in quantum mechanics, quantum optics and
quantum information\cite{Durt,Buks}. Solvable models in quantum
mechanics enable one to investigate quantum measurement problems
rather conveniently\cite{Nakazato,Shen1,Shen2}. A good number of
authors\cite{Stern,Aharonov,Srivastava,Sunn} have studied some
useful models such as Hepp-Coleman model \cite{Namiki} and Cini
model\cite{Cini}. The exact solvability of these models often
provides physicists with a clear understanding of the physical
phenomena involved and yield rich physical insights\cite
{Namiki2}. In this work, the first important step is to obtain the
exact solutions of the Schr\"{o}dinger equation and the
time-evolution operator that can be applied to the calculation of
the decoherence factor and study of the wavefunction collapse,
etc.. Although the exact solutions and the decoherence of these
models have been extensively investigated by many authors in the
literature\cite{Stern,Aharonov,Srivastava,Sunn,Namiki,Cini}, the
coefficients and parameters in these Hamiltonians are merely
time-independent ( or partially time-dependent ), to the best of
our knowledge. In the present letter, we obtain the explicit
time-evolution operator and the decoherence factor of the totally
time-dependent generalized Cini model, where all the parameters
depend on time. Since the time-dependent quantum model possesses
more rich properties, e. g., geometric phase and topological
feature\cite{Shen1,Shen2}, it is of essential significance to
consider the time-dependent case of quantum models.

Time-dependent system is governed by the time-dependent
Schr\"{o}dinger equation. The invariant theory\cite{Lewis}
suggested by Lewis and Riesenfeld in 1969 can solve the
time-dependent Schr\"{o}dinger equation. In 1991, Gao {\it et al.}
proposed a generalized invariant theory\cite{Gao0,Gao1}, by
introducing basic invariants, which enable one to find the
complete set of commuting invariants for some time-dependent
multi-dimensional systems\cite {Gao4,Shen,Kim}. Since the
time-dependent case and geometric phase factor has never been
considered in the time-dependent decoherence process, we will
analyze the generalized Cini model in what follows and then
calculate the time-dependent decoherence factor by making use of
these invariant theories.

The original Cini model for the correlation between the states of the
measured system and the measuring instrument-detector is built for a
two-level system interacting with the detector. Liu and Sun generalized this
Cini model to an $M$-level system\cite{Liu}. The Hamiltonian of this
generalized model is written

\begin{equation}
H=H_{S}+H_{D}+H_{I}, \eqnum{1} \label{eq1}
\end{equation}
where $H_{S}$ is the model Hamiltonian of the measured system S with $M$
levels and $H_{D}$ is the free Hamiltonian of the two-boson-state detector
D. They are generally of the forms

\begin{equation}
H_{S}=\sum_{k=1}^{M}E_{k}\left| \Phi _{k}\right\rangle
\left\langle \Phi _{k}\right| ,\quad H_{D}=\omega
_{1}a_{1}^{\dagger }a_{1}+\omega _{2}a_{2}^{\dagger }a_{2}
\eqnum{2} \label{eq2}
\end{equation}
with the creation and annihilation operators $a_{i}^{\dagger }$ , $a_{i}$
satisfying the following commuting relations

\begin{equation}
\left[ a_{i},a_{j}^{\dagger }\right] =\delta _{ij},\quad \left[ a_{i},a_{j}%
\right] =\left[ a_{i}^{\dagger },a_{j}^{\dagger }\right] =0.
\eqnum{3} \label{eq3}
\end{equation}
The interaction Hamiltonian $H_{I}$ is given by

\begin{equation}
H_{I}=\sum_{n}\left| \Phi _{n}\right\rangle \left\langle \Phi
_{n}\right| (g_{n}a_{1}^{\dagger }a_{2}+g_{n}^{\ast
}a_{2}^{\dagger }a_{1}). \eqnum{4} \label{eq4}
\end{equation}
In this letter, all the coefficients such as $E_{k},\omega
_{1},\omega _{2}$ and $g_{n}$ and $g_{n}^{\ast }$ in the
Hamiltonian are time-dependent, and the Schr\"{o}dinger equation
of the time-dependent generalized Cini model may be written

\begin{equation}
H(t)\left| \Psi (t)\right\rangle _{s}=i\frac{\partial }{\partial
t}\left| \Psi (t)\right\rangle _{s}. \eqnum{5} \label{eq5}
\end{equation}

It can be seen from the form of the Hamiltonian that both $\left| \Phi
_{k}\right\rangle \left\langle \Phi _{k}\right| $ and $N=\frac{%
a_{1}^{\dagger }a_{1}+a_{2}^{\dagger }a_{2}}{2}$ commute with $H,$ namely, $%
\left[ \left| \Phi _{k}\right\rangle \left\langle \Phi _{k}\right| ,H\right]
=\left[ N,H\right] =0.$ Hence, a generalized quasialgebra which enables one
to obtain the complete set of exact solutions of the Schr\"{o}dinger
equation can be found by working in a sub-Hilbert-space corresponding to the
particular eigenvalues of both $\left| \Phi _{k}\right\rangle \left\langle
\Phi _{k}\right| $ and $N.$ In order to use the Lewis-Riesenfeld invariant
theory\cite{Lewis}, we take $J_{+}=a_{1}^{\dagger
}a_{2},J_{-}=a_{2}^{\dagger }a_{1},J_{3}=\frac{a_{1}^{\dagger
}a_{1}-a_{2}^{\dagger }a_{2}}{2}$, which satisfy the commuting relations $%
[J_{3},J_{\pm }]=\pm J_{\pm },[J_{+},J_{-}]=2J_{3}.$ In this
sub-Hilbert-space the Hamiltonian can therefore be rewritten

\begin{equation}
H_{n,k}(t)=E_{k}+g_{k}J_{+}+g_{k}^{\ast }J_{-}+(\omega _{1}-\omega
_{2})J_{3}+n(\omega _{1}+\omega _{2}) \eqnum{6} \label{eq6}
\end{equation}
with $n$ being the eigenvalue of $N$ and satisfying

\begin{equation}
N\left| n_{1},n_{2}\right\rangle =n\left| n_{1},n_{2}\right\rangle ,\quad n=%
\frac{1}{2}(n_{1}+n_{2}). \eqnum{7}
\end{equation}
Thus in the sub-Hilbert-space we write the Schr\"{o}dinger equation in the
form

\begin{equation}
H_{n,k}(t)\left| \Psi _{n,k}(t)\right\rangle _{s}=i\frac{\partial
}{\partial t}\left| \Psi _{n,k}(t)\right\rangle _{s}, \eqnum{8}
\label{eq7}
\end{equation}
and $\left| \Psi (t)\right\rangle _{s}$ can be obtained from

\begin{equation}
\left| \Psi (t)\right\rangle _{s}=\sum_{n}\prod_{k}c_{n,k}\left|
\Psi _{n,k}(t)\right\rangle _{s}\left| \Phi _{k}\right\rangle,
\eqnum{9} \label{eq8}
\end{equation}
where $c_{n,k}$ is time-independent and determined by the initial conditions.

For the sake of using the invariant theory conveniently, we rewrite the
Hamiltonian $H_{n,k}(t)$ as follows

\begin{equation}
H_{n,k}(t)=c_{k}(t)\{\frac{1}{2}\sin \theta _{k}(t)\exp [-i\varphi
_{k}(t)]J_{+}+\frac{1}{2}\sin \theta _{k}(t)\exp [i\varphi
_{k}(t)]J_{-}+\cos \theta _{k}(t)J_{3}\}+f_{n,k}(t), \eqnum{10}
\label{eq9}
\end{equation}
where

\begin{equation}
c_{k}(t)\cos \theta _{k}(t)=\omega _{1}-\omega _{2},\quad
f_{n,k}(t)=E_{k}+n(\omega _{1}+\omega _{2}),\quad
\frac{1}{2}c_{k}(t)\sin \theta _{k}(t)\exp [-i\varphi
_{k}(t)]=g_{k}. \eqnum{11}
\end{equation}
In accordance with the invariant theory, an invariant that
satisfies the following invariant equation

\begin{equation}
\frac{\partial I_{k}(t)}{\partial
t}+\frac{1}{i}[I_{k}(t),H_{n,k}(t)]=0 \eqnum{12} \label{eq10}
\end{equation}
should be constructed. It follows from (\ref{eq10}) that the invariant $%
I_{k}(t)$ is the linear combination of $J_{\pm }$ and $J_{3}$ and may be
written as

\begin{equation}
I_{k}(t)=\frac{1}{2}\sin \lambda _{k}(t)\exp [-i\gamma _{k}(t)]J_{+}+\frac{1%
}{2}\sin \lambda _{k}(t)\exp [i\gamma _{k}(t)]J_{-}+\cos \lambda
_{k}(t)J_{3}. \eqnum{13} \label{eq11}
\end{equation}
Using the invariant equation (\ref{eq10}), two auxiliary equations by which $%
\lambda _{k}(t)$ and $\gamma _{k}(t)$ will be determined can be derived

\begin{equation}
\dot{\lambda}_{k}(t)=c_{k}\sin \theta _{k}\sin (\varphi
_{k}-\gamma _{k}),\quad \dot{\gamma}_{k}(t)=c_{k}[\cos \theta
_{k}-\sin \theta _{k}\cot \lambda _{k}\cos (\varphi _{k}-\gamma
_{k})].  \eqnum{14} \label{eq12}
\end{equation}

In order to obtain the exact solution of the time-dependent Schr\"{o}dinger
equation (\ref{eq7}), we introduce an invariant-related unitary
transformation operator $V_{k}(t)$

\begin{equation}
V_{k}(t)=\exp [\beta _{k}(t)J_{+}-\beta _{k}^{\ast }(t)J_{-}],
\eqnum{15} \label{eq13}
\end{equation}
where the time-dependent parameter

\begin{equation}
\beta _{k}(t)=-\frac{\lambda _{k}(t)}{2}\exp [-i\gamma
_{k}(t)],\quad \beta _{k}^{\ast }(t)=-\frac{\lambda
_{k}(t)}{2}\exp [i\gamma _{k}(t)]. \eqnum{16}\label{eq14}
\end{equation}
$V_{k}(t)$ can be readily shown to transform the time-dependent invariant $%
I_{k}(t)$ into $I_{kV}$ that is time-independent:

\begin{equation}
I_{kV}\equiv V_{k}^{\dagger }(t)I_{k}(t)V_{k}(t)=J_{3}. \eqnum{17}
\label{eq15}
\end{equation}
The eigenstate of the $I_{kV}=J_{3}$ corresponding to the eigenvalue $m$ is
denoted by $\left| j,m\right\rangle ,$ where

\begin{equation}
\left| j,m\right\rangle =\sum_{n_{1}n_{2}}\left\langle
n_{1},n_{2}\right| j,m\rangle \left| n_{1},n_{2}\right\rangle
,\quad m=\frac{1}{2}(n_{1}-n_{2}).\eqnum{18}
\end{equation}
By making use of $V_{k}(t)$ in expression (\ref{eq13}) and the
Baker-Campbell-Hausdorff formula\cite{Wei}

\begin{equation}
V^{\dagger }(t)\frac{\partial }{\partial t}V(t)=\frac{\partial }{\partial t}%
L+\frac{1}{2!}[\frac{\partial }{\partial t}L,L]+\frac{1}{3!}[[\frac{\partial
}{\partial t}L,L],L]+\frac{1}{4!}[[[\frac{\partial }{\partial t}%
L,L],L],L]+\cdots, \eqnum{19}
\end{equation}
where $V(t)=\exp [L(t)]$, one can obtain $H_{n,kV}(t)$ from $H_{n,k}(t)$

\begin{eqnarray}
H_{n,kV}(t) &=&V_{k}^{\dagger }(t)H_{n,k}(t)V_{k}(t)-V_{k}^{\dagger }(t)i%
\frac{\partial V_{k}(t)}{\partial t}  \nonumber \\
&=&\{[\cos \lambda _{k}\cos \theta _{k}+\sin \lambda _{k}\sin
\theta _{k}\cos (\gamma _{k}-\varphi
_{k})]+\dot{\gamma}_{k}(1-\cos \lambda _{k})\}J_{3}+f_{n,k}.
\eqnum{20} \label{eq16}
\end{eqnarray}
From the two expressions (\ref{eq15}) and (\ref{eq16}), one can see that $%
H_{n,kV}(t)$ differs from $I_{kV}$ only by a time-dependent
$c$-number factor and $f_{n,k}$. Use is made of the
invariant-related unitary transformation formulation and the
general solution of the time-dependent Schr\"{o}dinger equation
(\ref{eq7}) is therefore obtained

\begin{equation}
\left| \Psi _{n,k}(t)\right\rangle _{s}=%
\mathop{\textstyle\sum}%
_{m}C_{m}(n,k)\exp [i\phi _{m}(n,k,t)]V_{k}(t)\left|
j,m\right\rangle   \eqnum{21} \label{eq17}
\end{equation}
with the coefficients $C_{m}=\langle j,m,t=0\left| \Psi
_{n,k}(0)\right\rangle _{s}.$ The phase $\phi
_{m}(n,k,t)=\phi^{(d)} _{m}(n,k,t)+\phi^{(g)} _{m}(k,t)$ includes
the dynamical phase

\begin{eqnarray}
\phi^{(d)} _{m}(n,k,t) &=&-%
\textstyle\int%
_{0}^{t}\left\langle j,m\right| V_{k}^{\dagger }(t^{^{\prime
}})H_{n,k}(t^{^{\prime }})V_{k}(t^{^{\prime }})\left| j,m\right\rangle
{\rm {d}}t^{^{\prime }}  \nonumber \\
&=&-m%
\textstyle\int%
_{0}^{t}\{\cos \lambda _{k}(t^{^{\prime }})\cos \theta _{k}(t^{^{\prime }})+
\nonumber \\
&&+\sin \lambda _{k}(t^{^{\prime }})\sin \theta _{k}(t^{^{\prime }})\cos
[\gamma _{k}(t^{^{\prime }})-\varphi _{k}(t^{^{\prime }})]+\frac{1}{m}%
f_{n,k}(t^{^{\prime }})\}{\rm {d}}t^{^{\prime }} \eqnum{22}
\label{eq18}
\end{eqnarray}
and the geometric phase

\begin{equation}
\phi^{(g)} _{m}(k,t)=-\int_{0}^{t}\left\langle m\right|
-V_{k}^{\dagger }(t^{^{\prime }})i\frac{\partial V_{k}(t^{^{\prime
}})}{\partial t^{^{\prime
}}}\left| m\right\rangle {\rm {d}}t^{^{\prime }}=-m%
\textstyle\int%
_{0}^{t}\dot{\gamma}_{k}(t^{^{\prime }})[1-\cos \lambda
_{k}(t^{^{\prime }})]{\rm {d}}t^{^{\prime }}. \eqnum{23}
\label{eq19}
\end{equation}

It follows from (\ref {eq19}) that if the parameter
$\dot{\gamma}_{k}$ is taken to be time-independent, then the
geometric phase in one cycle ($T=\frac{2\pi}{\dot{\gamma}_{k}}$)
is $\phi^{(g)} _{m}(k,T)=-m[2\pi (1-\cos a)]$ where $2\pi (1-\cos
m)$ is an expression for the solid angle over the parameter space
of the invariant. It is of interest that $-m[2\pi (1-\cos a)]$ is
equal to the magnetic flux produced by a magnetic monopole ( and
the gravitomagnetic monopole ) of strength $\frac{\lambda }{4\pi
m}$ existing at the origin of the parameter space\cite{Gravshen}.
This, therefore, implies that geometric phase differs from
dynamical phase and it involves the global and topological
properties of the time evolution of a quantum system.

Since we have exact solutions of the time-dependent Cini model, we
can consider the decoherence factor that is given
\begin{equation}
F_{k,l}(t)=\left\langle j,m\right| V_{k}^{\dagger
}(t)V_{l}(t)\left| j,m\right\rangle . \eqnum{24} \label{eq20}
\end{equation}
Further calculation yields

\begin{equation}
F_{k,l}(t)=\exp [-m(\beta _{l}\beta _{k}^{\ast }-\beta _{k}\beta
_{l}^{\ast })]\left\langle j,m\right| \exp [(\beta _{l}-\beta
_{k})J_{+}-(\beta _{l}^{\ast }-\beta _{k}^{\ast })J_{-}]\left|
j,m\right\rangle,  \eqnum{25} \label{eq21}
\end{equation}
which is the general expression for the decoherence factor of the
time-dependent Cini model. Although the expression (\ref{eq21}) is
somewhat complicated, it is just the general explicit expression
that does not contain the chronological product. To show that the
decoherence factor (\ref{eq21}) can reduce to the familiar results
in the time-independent or partially time-dependent Cini model, we
consider a special and simple case where the Hamiltonian
is\cite{Zeng}

\begin{equation}
H_{n,k}=c_{k}J_{2}+f_{n,k}  \eqnum{26}\label{eq22}
\end{equation}
with $J_{2}=\frac{1}{2i}(J_{+}-J_{-}).$ It follows from
(\ref{eq22}) that, in this case, $\varphi _{k}=\frac{\pi
}{2},\omega _{1}=\omega _{2},\sin \theta _{k}=1.$ One can
therefore arrive at
\begin{equation}
\gamma _{k}=0,\quad \dot{\lambda}_{k}=c_{k},\quad \beta _{k}=\beta
_{k}^{\ast }=-\frac{\lambda _{k}}{2}.  \eqnum{27}\label{q23}
\end{equation}
In the similar fashion, we have

\begin{equation}
\gamma _{l}=0,\quad \dot{\lambda}_{l}=c_{l},\quad \beta _{l}=\beta
_{l}^{\ast }=-\frac{\lambda _{l}}{2}.  \eqnum{28}\label{q24}
\end{equation}
It is verified with the help of (\ref{eq21}) that

\begin{equation}
F_{k,l}(t)=\left\langle j,m\right| \exp
[i\int_{0}^{t}(c_{k}-c_{l}){\rm {d}}t^{^{\prime }}J_{2}]\left|
j,m\right\rangle . \eqnum{29}\label{q25}
\end{equation}
If the state of the measuring instrument-detector at the initial
$t=0$ is $\left| j,j\right\rangle $, then the decoherence factor
is

\begin{eqnarray}
F_{k,l}(t) &=&\left\langle j,j\right| \exp
[i\int_{0}^{t}(c_{k}-c_{l}){\rm {d}}t^{^{\prime }}J_{2}]\left|
j,j\right\rangle
\nonumber \\
&=&[\cos (\int_{0}^{t}\frac{c_{k}-c_{l}}{2}{\rm {d}}t^{^{\prime
}})]^{2j}. \eqnum{30}\label{q26}
\end{eqnarray}
Note that the expression (\ref{q26}) is a well-known result and in
complete agreement
with the one obtained by Sun {\it et al.}\cite{Zeng}. When $j\rightarrow \infty $ and $%
\int_{0}^{t}\frac{c_{k}-c_{l}}{2}{\rm {d}}t^{^{\prime }}\neq n\pi
\quad (n=0,\pm 1,\pm 2,\cdots ),\quad F_{k,l}(t)\rightarrow 0$,
which means the wavefunction collapse occurs under the classical
limit.

To conclude this letter, we briefly discuss the concepts of the
exact solution and the explicit solution. The expression
(\ref{eq17}) is a particular exact solution corresponding to the
particular eigenvalue $m$ of the invariant and thus the general
solutions of the time-dependent Schr\"{o}dinger equation are
easily obtained by using the linear combinations of all these
particular solutions. Generally speaking, in Quantum Mechanics,
solution with chronological-product operator ( time-order operator
) $P$ is often called the formal solution. In the present letter,
however, the solution of the Schr\"{o}dinger equation governing a
time-dependent system is sometimes called the explicit solution,
for reasons that the solution does not involve time-order
operator. But, on the other hand, by using Lewis-Riesenfeld
invariant theory, there always exist time-dependent parameters,
for instance, $\lambda_{k}(t)$ and $\gamma_{k}(t)$ in this letter,
which are determined by the auxiliary equations (\ref{eq12}).
According to the traditional practice, when employed in
experimental analysis and compared with experimental results,
these nonlinear auxiliary equations should be solved often by
means of numerical calculation. From above viewpoints, the concept
of explicit solution is understood in a relative sense, namely, it
can be
considered explicit solution when compared with the time-evolution operator $%
U(t)=P\exp [\frac{1}{i}\int_{0}^{t}H(t^{^{\prime }}){\rm
d}t^{^{\prime }}]$ involving time-order operator, $P$; whereas, it
cannot be considered completely explicit solution for it is
expressed in terms of some time-dependent parameters which should
be obtained via the auxiliary equations. Hence, conservatively
speaking, we regard the solution of the time-dependent system
presented in the letter as exact solution rather than explicit
solution.

The present letter obtains exact solutions and decoherence factor
of the time-dependent Cini model by working in the sub-Hilbert
space corresponding to the eigenvalue of two invariants, which
commute with the Hamiltonian, and by using the invariant-related
unitary transformation method. The invariant-related unitary
transformation formulation is an effective method for treating
time-dependent problems. This formulation replaces eigenstates of
the time-dependent invariants with those of the time-independent
invariants through the unitary transformation. It uses the
invariant-related unitary transformation and obtains the explicit
expression for the time-evolution operator, instead of the formal
solution associated with the chronological product. In view of
what has been discussed above, it can be seen that the invariant
theory is appropriate to treat the time-dependent quantum
decoherence. Apparently, this method is easy to generalize to the
time-dependent Hepp-Coleman model\cite{Namiki}. Since the
geometric phase factor appears in the time-dependent systems, it
is interesting to consider the geometric phase in the
time-dependent quantum decoherence.

Acknowledgements The authors thank Gao Xiao-Chun for his
beneficial invariant-related unitary transformation formulation.

\end{document}